\documentclass[twoside,twocolumn,9pt]{article}
\usepackage{extsizes}
\usepackage[super,sort&compress,comma]{natbib}
\usepackage[version=3]{mhchem}
\usepackage[left=1.5cm, right=1.5cm, top=1.785cm, bottom=2.0cm]{geometry}
\usepackage{balance}

\usepackage{sectsty}
\usepackage{graphicx}
\usepackage[format=plain,justification=justified,singlelinecheck=false,font={stretch=1.125,small,sf},labelfont=bf,labelsep=space]{caption}
\usepackage{float}
\usepackage{fancyhdr}
\usepackage{fnpos}
\usepackage[english]{babel}
\addto{\captionsenglish}{%
  
}
\usepackage{array}
\usepackage{droidsans}
\usepackage{charter}
\usepackage[T1]{fontenc}
\usepackage[usenames,dvipsnames]{xcolor}
\usepackage{setspace}
\usepackage[compact]{titlesec}
\usepackage{hyperref}
\usepackage{amsmath}
\usepackage{amssymb}
\usepackage{bm}
\usepackage{subcaption}

\usepackage{epstopdf}
\usepackage{siunitx}

\definecolor{cream}{RGB}{222,217,201}

\usepackage[normalem]{ulem}
\usepackage[textwidth=\dimexpr\textwidth-2cm\relax]{todonotes}
\makeatletter
\@mparswitchfalse%
\makeatother
\normalmarginpar 

\usepackage[T1]{fontenc}
\usepackage[utf8]{inputenc}
\usepackage{textcomp}
\usepackage{helvet}
\usepackage{microtype}

\usepackage{hyperref}
\usepackage{lastpage}

\begin{document}

\pagestyle{fancy}
\thispagestyle{plain}
\fancypagestyle{plain}{
\renewcommand{\headrulewidth}{0pt}
}

\makeFNbottom
\makeatletter
\renewcommand\LARGE{\@setfontsize\LARGE{15pt}{17}}
\renewcommand\Large{\@setfontsize\Large{12pt}{14}}
\renewcommand\large{\@setfontsize\large{10pt}{12}}
\renewcommand\footnotesize{\@setfontsize\footnotesize{7pt}{10}}
\makeatother

\renewcommand{\thefootnote}{\fnsymbol{footnote}}
\renewcommand\footnoterule{\vspace*{1pt}%
\color{cream}\hrule width 3.5in height 0.4pt \color{black}\vspace*{5pt}}
\setcounter{secnumdepth}{5}

\makeatletter
\renewcommand\@biblabel[1]{#1}
\renewcommand\@makefntext[1]%
{\noindent\makebox[0pt][r]{\@thefnmark\,}#1}
\makeatother
\renewcommand{\figurename}{\small{Fig.}~}
\sectionfont{\sffamily\Large}
\subsectionfont{\normalsize}
\subsubsectionfont{\bf}
\setstretch{1.125} 
\setlength{\skip\footins}{0.8cm}
\setlength{\footnotesep}{0.25cm}
\setlength{\jot}{10pt}
\titlespacing*{\section}{0pt}{4pt}{4pt}
\titlespacing*{\subsection}{0pt}{15pt}{1pt}

\fancyfoot{}
\fancyfoot[LO,RE]{\vspace{-7.1pt}\includegraphics[height=9pt]{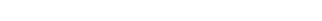}}
\fancyfoot[CO]{\vspace{-7.1pt}\hspace{13.2cm}\includegraphics{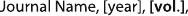}}
\fancyfoot[CE]{\vspace{-7.2pt}\hspace{-14.2cm}\includegraphics{head_foot/RF}}
\fancyfoot[RO]{\footnotesize{\sffamily{1--\pageref{LastPage} ~\textbar  \hspace{2pt}\thepage}}}
\fancyfoot[LE]{\footnotesize{\sffamily{\thepage~\textbar\hspace{3.45cm} 1--\pageref{LastPage}}}}
\fancyhead{}
\renewcommand{\headrulewidth}{0pt}
\renewcommand{\footrulewidth}{0pt}
\setlength{\arrayrulewidth}{1pt}
\setlength{\columnsep}{6.5mm}
\setlength\bibsep{1pt}

\makeatletter
\newlength{\figrulesep}
\setlength{\figrulesep}{0.5\textfloatsep}

\newcommand{\topfigrule}{\vspace*{-1pt}%
\noindent{\color{cream}\rule[-\figrulesep]{\columnwidth}{1.5pt}} }

\newcommand{\botfigrule}{\vspace*{-2pt}%
\noindent{\color{cream}\rule[\figrulesep]{\columnwidth}{1.5pt}} }

\newcommand{\dblfigrule}{\vspace*{-1pt}%
\noindent{\color{cream}\rule[-\figrulesep]{\textwidth}{1.5pt}} }

\makeatother

\twocolumn[
  \begin{@twocolumnfalse}
{\includegraphics[height=30pt]{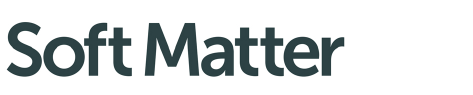}\hfill\raisebox{0pt}[0pt][0pt]{\includegraphics[height=55pt]{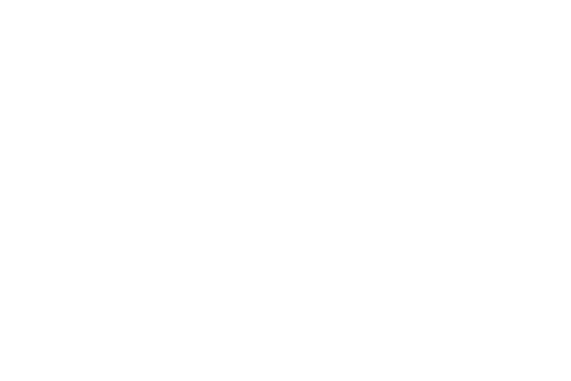}}\\[1ex]
\includegraphics[width=18.5cm]{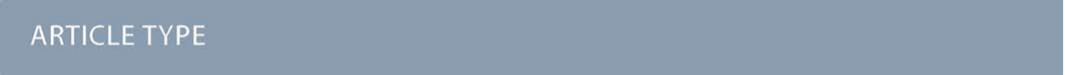}}\par
\vspace{1em}
\sffamily
\begin{tabular}{m{4.5cm} p{13.5cm} }

\includegraphics{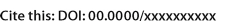} & \noindent\LARGE{\textbf{Impacting spheres:
from liquid drops to elastic beads$^\dag$}} \\
\vspace{0.3cm} & \vspace{0.3cm} \\

& \noindent\large{Saumili Jana\textit{$^{a,1}$}, John Kolinski\textit{$^{b,2}$}, Detlef
Lohse\textit{$^{a,c,3}$}, and Vatsal Sanjay\textit{$^{e,4}$}} \\

\includegraphics{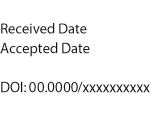} & \noindent\normalsize{
  A liquid drop impacting a non‑wetting rigid substrate spreads laterally,
  then retracts, and finally jumps off again. An elastic solid, by contrast, undergoes a slight
  deformation, contacts briefly, and bounces. The impact force on the substrate -- crucial
  for engineering and natural processes -- is classically described by Wagner’s (liquids)
  and Hertz’s (solids) theories. This work bridges these limits by considering a generic
  viscoelastic medium. Using direct numerical simulations, we study a viscoelastic sphere
  impacting a rigid, non‑contacting surface and quantify how the elasticity number
  ($El$, dimensionless elastic modulus) and the Weissenberg number ($Wi$, dimensionless
  relaxation time) dictate the impact force. We recover the Newtonian liquid response as
  either $El \to 0$ or $Wi \to 0$, and obtain elastic‑solid behavior in the limit $Wi \to
  \infty$ and $El \ne 0$. In this elastic‑memory limit, three regimes
  emerge -- capillary‑dominated, Wagner scaling, and Hertz scaling -- with a smooth transition
  from the Wagner to the Hertz regime. Sweeping $Wi$ from 0 to $\infty$ reveals a
  continuous shift from materials with no memory to materials with permanent memory of
  deformation, providing an alternate, controlled route from liquid drops to elastic
  beads. The study unifies liquid and solid impact processes and offers a general
  framework for the liquid‑to‑elastic transition relevant across systems and applications.
} \\

\end{tabular}

 \end{@twocolumnfalse} \vspace{0.6cm}

  ]

\renewcommand*\rmdefault{bch}\normalfont\upshape
\rmfamily
\section*{}
\vspace{-1cm}


\footnotetext{\textit{$^{a}$~Physics of Fluids Department, Max Planck Center Twente for Complex Fluid Dynamics, and J. M. Burgers Center for Fluid Dynamics, University of Twente, P.O. Box 217, 7500AE Enschede, the Netherlands.}}
\footnotetext{\textit{$^{b}$~Institute of Mechanical Engineering, School of Engineering, EPFL, Lausanne, Switzerland.}}
\footnotetext{\textit{$^{c}$~Max Planck Institute for Dynamics and Self-Organisation, Am Fassberg 17, 37077 Göttingen, Germany.}}
\footnotetext{\textit{$^{d}$~CoMPhy Lab, Department of Physics, Durham University, Science Laboratories, South Road, Durham DH1 3LE, United Kingdom.}}

\footnotetext{$^{1}$~s.jana@utwente.nl}
\footnotetext{$^{2}$~john.kolinski@epfl.ch}
\footnotetext{$^{3}$~d.lohse@utwente.nl}
\footnotetext{$^{4}$~vatsal.sanjay@comphy-lab.org}



\section{Introduction}\label{introduction}
Impacts of spherical bodies on rigid substrates span two classical limits that have long
been treated separately: liquid drops \citep{wagner1932stoss,
Philippi2016,zhangImpactForcesWater2022,sanjayRoleViscosityDrop2025,sanjayUnifyingTheoryScaling2025} and
elastic solids
\citep{hertzUeberBeruhrungFester1882,pao1955extension,hunterEnergyAbsorbedElastic1957,bertinSimilaritySolutionsElastohydrodynamic2024}.
Both occur widely in nature and technology, where the normal force on the substrate is
often the quantity of interest because it can damage engineered surfaces
\citep{turbine_blade_wear2013, cheng2021drop}. Drop impact governs processes from inkjet
printing \citep{lohse2022fundamental} and spray cooling/coating \citep{cooling_2017,
spray_cool_Kim_2007} to forensics \citep{forensics_2018}, pesticide deposition
\citep{pesticide}, and soil erosion \citep{Nearing1986}. Impacts of elastic solids arise
in hardness testing \citep{tirupataiah1991}, granular media and suspensions
\citep{granularmedia_Andreotti_2013}, sports \citep{sports}, and everyday bouncing of soft
rubber balls. Despite this breadth, a unifying framework for the impact force across
liquid and solid limits has remained elusive.

A falling liquid drop, after impact on a rigid surface deforms and spreads laterally
until it reaches its maximum extent. A pronounced peak in the temporal evolution of force occurs
at the instance of drop touchdown on the surface due to the inertia of the impact, whereas during
droplet spreading this force is much smaller \citep{sanjayUnifyingTheoryScaling2025}. For perfectly
wetting surfaces, the liquid sticks to it.
However, for non-wetting surfaces, the drop retracts from its maximum
spread and generates a Worthington jet which coincides with a second peak in the temporal
evolution of force \citep{zhangImpactForcesWater2022,sanjayRoleViscosityDrop2025}.
In the inertial regime, Wagner's theory predicts that the impact force scales as
\begin{equation}
    F \sim \rho_l V_0^2 R_0^2,
    \label{wagner_scale}
\end{equation}
where $\rho_l$ is the density of the media, $V_0$ is the impact velocity and $R_0$ is the
radius of the falling drop \citep{cheng2021drop}.

By contrast, the elastic solids undergo slight deformation on impact with the substrate
and bounce off following a brief contact with the substrate due to the exerted normal
reaction. For such cases, the temporal evolution of force is characterized by a single
maximum. Assuming the contact area to be small in comparison to the bead's size, and
considering a non-adhesive contact with small strains within the elastic limit, the
situation can be treated as a Hertzian contact problem
\citep{bertinSimilaritySolutionsElastohydrodynamic2024}. Thus, Hertz's theory describes the
scaling laws for the impact force in the case of the impact of an elastic bead on a rigid
substrate,
\begin{equation}
    F \sim (GR_0^2)^{2/5}(\rho V_0^2 R_0^2)^{3/5}
    \label{hertztheory}
\end{equation}
where $G$ is the modulus of rigidity, $\rho$ is the density of the solid medium, and $R_0$
and $V_0$ as mentioned before are the radius and impact velocity of the solid elastic
bead.

Viscoelastic media -- here, soft elastic gels -- bridge liquids and
solids: when deformed they support both viscous flow and recoverable elastic stress
\citep{Snoeijer_etal_2020}. Compared with Newtonian liquids, their rate‑dependent rheology
can markedly alter spreading, pinch‑off, and rebound on impact
\citep{clasen2006jfm,viscoelastic_drp_oscil_2021}. Such soft media are relevant to inkjet
printing \citep{inkjet_materials_Calvert}, drop deposition
\citep{control_deposition_visc}, and spray atomization \citep{spray_atomisation_visc}.
Soft solids such as hydrogels, comprising cross‑linked networks with tunable elasticity,
are widely used as biocompatible materials in rapid prototyping
\citep{prototype_derby_science} and drug delivery \citep{drug_delivery}; see
\citet{soft_materials_annrev} for background.

In this work we parameterize the gel’s elastic response using the elastic modulus $G$
which is the proportionality constant between strain and elastic stresses, and the
relaxation time $\lambda$ that sets the decay timescale of those elastic stresses. Upon
non-dimensionalizing the governing equations (section \ref{subsec:gov_eqns}), two material
control parameters emerge. The elastocapillary number
\begin{equation}
    Ec = \frac{GR_0}{\gamma}
    \label{eq:Ec}
\end{equation}
compares the elastic modulus to the Laplace pressure, while the Deborah number
\begin{equation}
    De = \frac{\lambda}{\sqrt{\rho_l R_0^3/\gamma}}
    \label{eq:De}
\end{equation}
compares the elastic-stress relaxation timescale to the inertio-capillary process time. Here $\gamma$ represents the coefficient of surface tension.

The impact inertia is expressed by the Weber number
\begin{equation}
    We  = \frac{\rho_l V_0^2 R_0}{\gamma},
    \label{eq:We}
\end{equation}
which compares the inertial and capillary forces.
Another important dimensionless control parameter of the system is the Ohnesorge number
\begin{equation}
    Oh = \frac{\eta_l}{\sqrt{\rho_l\gamma R_0}}
    \label{eq:Oh}
\end{equation}
which is the ratio of inertio-capillary and the visco-capillary timescales. Here $\eta_l$ is the dynamic viscosity of the sphere.

Two combinations of these numbers will be central in what follows. The elasticity number
\begin{equation}
El = \frac{Ec}{We} = \frac{G}{\rho_l V_0^2},
\label{eq:El}
\end{equation}
compares elastic to inertial stresses. The Weissenberg number
\begin{equation}
Wi = De \sqrt{We} = \frac{\lambda}{R_0/V_0},
\label{eq:Wi}
\end{equation}
compares the elastic relaxation time $\lambda$ to the impact time $R_0/V_0$ (see
\S~\ref{subsec:gov_eqns}).


In this study, we simulate impacts of soft gel spheres on a rigid, non‑contacting
substrate using a volume‑of‑fluid, finite‑volume framework. By varying the elastocapillary
number ($Ec$) and the Deborah number ($De$), we traverse smoothly from liquid‑like to
solid‑like response and compare the resulting force scalings with Wagner’s (liquids) and
Hertz’s (elastic solids) theories. We develop an expression for the peak force that transcends
the two regimes, using a function of the elasticity parameter to compare the shear modulus with
the impact stress. Consistent with these limits, we recover
Newtonian‑liquid behavior for $De=0$ or $Ec=0$, while for $De\to\infty$ at sufficiently
large $Ec$ the dynamics converge to those of an elastic solid.

\section{Numerical Framework} \label{sec:numerics}
\subsection{Problem description and governing equations}  \label{subsec:gov_eqns}
\begin{figure}[h]
\centering
  \includegraphics[height=4cm]{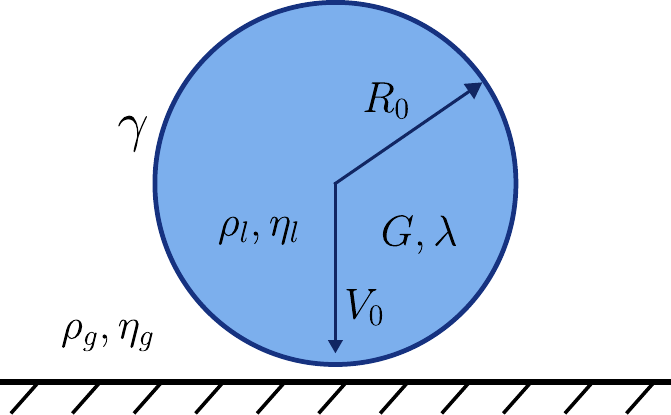}
  \caption{Schematic: a viscoelastic sphere (radius $R_0$) impacts a non‑contacting rigid
  surface with velocity $V_0$. Material properties are $\rho_l$, $\eta_l$, $G$, $\lambda$,
and $\gamma$ for the sphere; $\rho_g$ and $\eta_g$ for the gas.}
  \label{fgr:schematic}
\end{figure}

We consider an axisymmetric sphere of radius $R_0$ approaching a rigid substrate with
initial velocity $V_0$. For liquid drops the substrate is non‑wetting; for viscoelastic
and elastic beads it is non‑contacting. The sphere is a viscoelastic medium of density
$\rho_l$, dynamic viscosity $\eta_l$, elastic modulus $G$, relaxation time $\lambda$, and
surface tension coefficient $\gamma$. The surrounding gas has density $\rho_g$ and
viscosity $\eta_g$ (figure~\ref{fgr:schematic}).

Lengths are scaled by $R_0$ and time by the inertio‑capillary timescale
$\tau_\gamma=\sqrt{\rho_l R_0^3/\gamma}$. The corresponding velocity and pressure/stress
scales are $u_\gamma=\sqrt{\gamma/(\rho_l R_0)}$ and $\sigma_\gamma=\gamma/R_0$,
respectively. Throughout the manuscript, all variables with tilde are non-dimensionalized using
the above mentioned scales. The incompressible mass and
momentum balances in the viscoelastic phase read

\begin{equation}
    \boldsymbol{\nabla\cdot}\bm{u} = 0
    \label{eq:mass_cons}
\end{equation}
and,
\begin{equation}
   \frac{\partial(\rho\bm{u})}{\partial t} + \boldsymbol{\nabla\cdot}(\rho\bm{uu})
   = -\boldsymbol{\nabla}p +
   \boldsymbol{\nabla\cdot}(\boldsymbol{\sigma_v}+\boldsymbol{\sigma_e}) +
   \boldsymbol{f}_\gamma
   \label{eq:mom_cons}
\end{equation}
where $\boldsymbol{f}_\gamma$ is the non‑dimensional capillary force density acting at the
interface. The Newtonian (viscous) stress is
\begin{equation}
  \boldsymbol{\tilde{\sigma_v}} = 2Oh\boldsymbol{\tilde{\mathcal{D}}}
    \label{eq:sigma_v}
\end{equation}
with $\boldsymbol{\tilde{\mathcal{D}}} = (\boldsymbol{\tilde{\nabla}}\bm{\tilde{u}} +
\boldsymbol{\tilde{\nabla}}\bm{\tilde{u}}^T)/2$ denoting the symmetric part of the velocity gradient
tensor.
Further, the normal force on the substrate is obtained using the rate of change of sphere's momentum

\begin{equation}
     \boldsymbol{F}(t) = \frac{4}{3}\pi R_0^3 \rho_l \frac{dV_{cm}}{dt},
     \label{eq:force_vcm}
\end{equation}
 where $V_{cm}$ denotes the velocity of the centre of mass of the drop at any instant.

The elastic stresses arise from deformation of the microstructure quantified by the
conformation tension $\boldsymbol{\mathcal{A}}$ \citep{Snoeijer_etal_2020}.
Using the Oldroyd‑B constitutive model \citep{Oldroyd_1950},

\begin{equation}
  \boldsymbol{\tilde{\sigma_e}} = Ec(\boldsymbol{\mathcal{A}} - \boldsymbol{\mathcal{I}})
    \label{eq:sima_p}
\end{equation}
with the elastocapillary number $Ec=GR_0/\gamma$ (eq.~\ref{eq:Ec}).
The conformation tensor relaxes to $\boldsymbol{\mathcal{I}}$ on the Deborah
timescale $De=\lambda/\tau_\gamma$ (eq.~\ref{eq:De}) via

\begin{equation}
  \stackrel{\smash{\raisebox{0ex}{$\mkern8mu\boldsymbol{\nabla}$}}}{\boldsymbol{\mathcal{A}}}
  = -\frac{1}{De}(\boldsymbol{\mathcal{A}}-\boldsymbol{\mathcal{I}}),
\label{eq:relax_law}
\end{equation}
where the upper‑convected derivative is

\begin{equation}
  \stackrel{\smash{\raisebox{0ex}{$\mkern8mu\boldsymbol{\nabla}$}}}{\boldsymbol{\mathcal{A}}}
  \equiv \frac{\partial \boldsymbol{\mathcal{A}}}{\partial \tilde{t}} + (\bm{\tilde{u}}\cdot
  \boldsymbol{\tilde{\nabla}})\boldsymbol{\mathcal{A}} -
  2\text{Sym}(\boldsymbol{\mathcal{A}}\cdot(\boldsymbol{\tilde{\nabla} \tilde u})).
\end{equation}

The Deborah number $De$ quantifies material memory: $De=0$ recovers a Newtonian liquid
characterized by $Oh$; $De\to\infty$ yields an elastic solid limit with

\begin{equation}
  \stackrel{\smash{\raisebox{0ex}{$\mkern8mu\boldsymbol{\nabla}$}}}{\boldsymbol{\mathcal{A}}}=0
\end{equation}
and Oldroyd‑B equivalent to a neo‑Hookean solid \citep{Snoeijer_etal_2020}.

Despite being widespread due to simplicity, the Oldroyd-B model suffers from certain
limitations \citep{Snoeijer_etal_2020,dixitViscoelasticWorthingtonJets2025}.
It fails to capture the shear-thinning behaviour
in viscoelastic fluids completely, and erroneously predicts unbounded stress growth in strong
extensional flows \citep{Mckinley_exten_flow}. This limitation can be addressed by
incorporating the finite polymer extension (FENE-P model)\citep{FENE-P_original}. Also various other extensions of the Oldroyd-B model have
been developed \citep{tanner2000} like the Phan-Thien-Tanner (PTT)\citep{Thien_1977} model
to account for such non-linearities. However, since we are not dealing with strong
extensional flows, we restrict ourselves to the Oldroyd-B model in this study, as
it is sufficient to describe our case.

\subsection{Numerical methods and simulations} \label{subsec:methods}

We solve the above equations with \textsc{Basilisk} C \citep{popinet-basilisk}, using a
one‑fluid formulation with surface tension as a singular interfacial force
\citep{tryggvason2011direct,brackbill1992continuum}. The liquid–gas interface is tracked
by a volume‑of‑fluid (VoF) color function $\psi$ advected by

\begin{equation}
  \frac{\partial \psi}{\partial t}+\boldsymbol{\nabla}\cdot(\psi,\bm{u})=0,
\label{eq:vof}
\end{equation}
with $\psi=1$ in liquid, $\psi=0$ in gas, and $0<\psi<1$ at the interface. Mixture properties are

\begin{equation}
  \rho = \psi\rho_l+(1-\psi)\rho_g,
\label{eq:density}
\end{equation}

\begin{equation}
  \eta = \psi\eta_l+(1-\psi)\eta_g,
\label{eq:viscosity}
\end{equation}
with fixed ratios $\rho_r=\rho_g/\rho_l=10^{-3}$ and $\eta_r=\eta_g/\eta_l=10^{-2}$  and a small sphere Ohnesorge number, $Oh = 10^{-2}$.
A geometric VoF reconstruction applies capillary forces as

\begin{equation}
  \boldsymbol{f}_\gamma \approx \gamma\kappa\boldsymbol{\nabla}\psi,
\label{eq:surface_tension}
\end{equation}
where the curvature $\kappa$ is computed via height functions
\citep{popinet2018numerical}. Explicit surface‑tension forcing imposes the standard
capillary time‑step constraint \citep{popinet2009}; the explicit update of
$\boldsymbol{\sigma_e}$ adds a typically milder constraint.

At the substrate we impose no‑penetration and no‑slip, and a zero normal pressure
gradient. To enforce a non‑contacting (superhydrophobic) condition we set $\psi=0$ at the
wall, maintaining a thin air cushion \citep{kolinski-2014-epl,Vatsal_thesis}. For liquid drops, enforcing this air cushion results in a non-wetting (superhydrophobic) substrate, while for elastic spheres this results in a non-contacting substrate. We stress
that contact initiation in soft–solid impacts is generically air‑mediated and can proceed
annularly or patchily with a non‑monotonic initial contact radius; see Zheng et al.
(2021)\citep{zheng2021air} for direct observations of air-mediated contact in
compliant‑hemisphere impacts. Top and lateral boundaries use outflow (ambient pressure,
zero tangential stress, zero normal velocity gradient). Boundaries are positioned far
enough to avoid spurious confinement effects. The axisymmetric domain size is $8R_0\times
8R_0$. We employ quadtree adaptive mesh refinement (AMR) \citep{popinet2009,popinet2015}
with maximal refinement at the interface and in regions of large velocity gradients.
Wavelet‑based error control uses tolerances $10^{-3}$ for $\bm{u}$, $\psi$, $\kappa$, and
$\boldsymbol{\mathcal{A}}$. Grid‑independence tests confirm convergence. Unless stated
otherwise, the minimum cell size is $\Delta=R_0/512$ (i.e.\ 512 cells per radius on a
uniform equivalent grid), increased to $\Delta=R_0/2048$ when required. Further numerical
details can be found in
\citet{popinet2015,Vatsal_thesis} and \citet{dixitViscoelasticWorthingtonJets2025}.

\section{Wagner versus Hertz: Permanent‑Memory Impacts} \label{infDe}

In this section, we quantify the solid‑impact limit by taking $De\to\infty$, so the
material retains its deformation memory over the process time. In the numerics, we keep a
small background viscosity, so the spheres are Kelvin–Voigt solids rather than
perfectly elastic; this facilitates comparison with inertial liquid impacts at finite
$Oh$ and avoids the numerical breakdown of the inviscid (Euler)–elastic limit. We
therefore approach the purely elastic response by letting $Oh\to 0$.

We sweep the $(Ec,We)$ space over $We\in[1,10^{3}]$ and $Ec\in[10^{-1},10^{4}]$. The normal reaction on the substrate is
computed from the drop’s momentum balance (eq.~\ref{eq:force_vcm}). For liquid drops on non-wetting substrates, $F(t)$ exhibits two peaks: an inertial peak at touchdown and a later peak associated with the formation of a Worthington jet \citep{zhangImpactForcesWater2022}. The second, jetting peak occurs at a time $t_2 \gg t_{\max}$, with the ratio $t_2/t_{\max} \sim \sqrt{We}$ \citep{sanjayRoleViscosityDrop2025}, reflecting the phase difference between recoil and jet formation. In contrast, for elastic spheres the loading and unloading remain nearly in phase: the entire contact–rebound cycle fits within $t \lesssim 2t_{\max}$ and $F(t)$ displays only a single, almost symmetric peak in this interval. To compare liquids and solids and to track the transition, we therefore focus on the first (inertial) peak $F_{\max}$, non-dimensionalized as $F_{\max}/(\rho_l V_0^2 R_0^2)$; the corresponding time is $t_{\max}$.

\begin{figure*}
  \centering
  \includegraphics[width=0.98\textwidth]{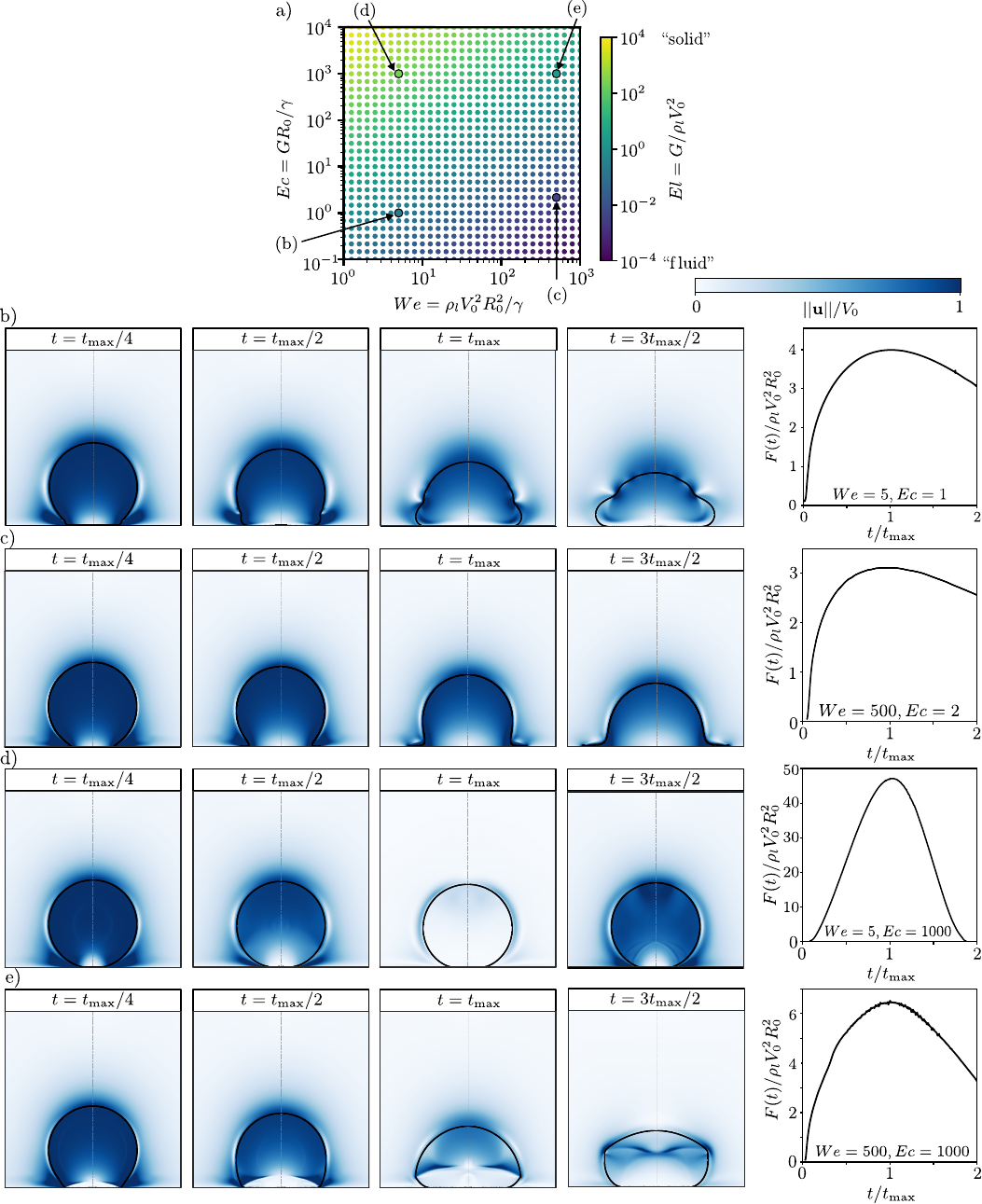}
  \caption{(a) Phase space in the $Ec$-$We$ plane illustrating the range of simulations
    conducted in this work colored according to the elasticity number $El = Ec/We$. The
    four highlighted symbols locate typical cases representing the range of parameters
    explored. We chose ($We, Ec$) $=$ (b) ($5,1$), (c) ($500,2$), (d) ($5, 1000$), (e)
    ($500, 1000$). For each case, the color scheme of each snapshot represents the
  magnitude of the velocity normalized by the impact velocity, alongside the corresponding force history $F(t)/(\rho_l V_0^2 R_0^2)$ plotted versus $t/t_{\max}$ (right). The force traces are plotted up to $t/t_{\max}=2$: for the liquid-drop reference, the second peak associated with the Worthington jet \cite{zhangImpactForcesWater2022,sanjayRoleViscosityDrop2025} occurs at later times $t_2 \gg t_{\max}$ (with $t_2/t_{\max} \sim \sqrt{We}$) and is therefore outside the plotted window.}
  \label{fgr:diff_regimes}
\end{figure*}

Figure~\ref{fgr:diff_regimes} shows representative cases across the parameter space. At
low $Ec$ the sphere flows and behaves liquid‑like (figs.~\ref{fgr:diff_regimes}b,c;
$Ec=1,2$). At low $Ec$ and low $We$, capillarity is significant
(fig.~\ref{fgr:diff_regimes}b). At high $Ec$ the sphere deforms slightly and rebounds
after a short contact (figs.~\ref{fgr:diff_regimes}d; $Ec=1000$). Increasing $We$ at fixed
$Ec$ effectively softens the response (reduces $El$) and increases the contact duration
(figs.~\ref{fgr:diff_regimes}d). The force traces reflect this evolution: for large $El$
(high $Ec$, low $We$) $F(t)$ is nearly symmetric, as in elastic impacts, while decreasing
$El$ (e.g. by increasing $We$) skews $F(t)$ in the manner typical of liquid impacts
\citep{zhangImpactForcesWater2022,sanjayRoleViscosityDrop2025}. The peak magnitude also
varies appreciably across cases.
Since force and time are related as $F_{max}/(\rho_lV_0^2 R_0^2)\cdot t_{max}/(R_0/V_0) \sim 1$ for any impacting sphere in general, $t_{max}$ reduces accordingly with an increase in $F_{max}$.

The dependence of $F_{\max}$ on $We$ and $Ec$ is summarized in figure~\ref{fgr:fvswe,ec}.
For $Ec\lesssim1$, $F_{\max}$ follows the liquid‑impact trend with the low‑$We$ correction
\citep{zhangImpactForcesWater2022},

\begin{equation}
  \frac{F_{\max}}{\rho_l V_0^2 R_0^2}\approx \frac{3.2}{We}+3.24.
  \label{eq:fmax_liq}
\end{equation}

As $Ec$ increases, $F_{\max}$ rises, most strongly at low $We$. At sufficiently large
$Ec$, $F_{\max}/(\rho_l V_0^2 R_0^2)$ decreases with $We$ with a log–log slope
$\simeq-2/5$, indicating a transition from eq.~\eqref{eq:fmax_liq} to $F_{\max}/(\rho_l
V_0^2 R_0^2)\sim We^{-2/5}$. At fixed $We$ (figure~\ref{fgr:fvswe,ec}b), $F_{\max}$ is
nearly constant at small $Ec$ and then increases steadily with $Ec$, with higher
magnitudes at lower $We$.

\begin{figure*}
  \centering
  \includegraphics[width=0.9\textwidth]{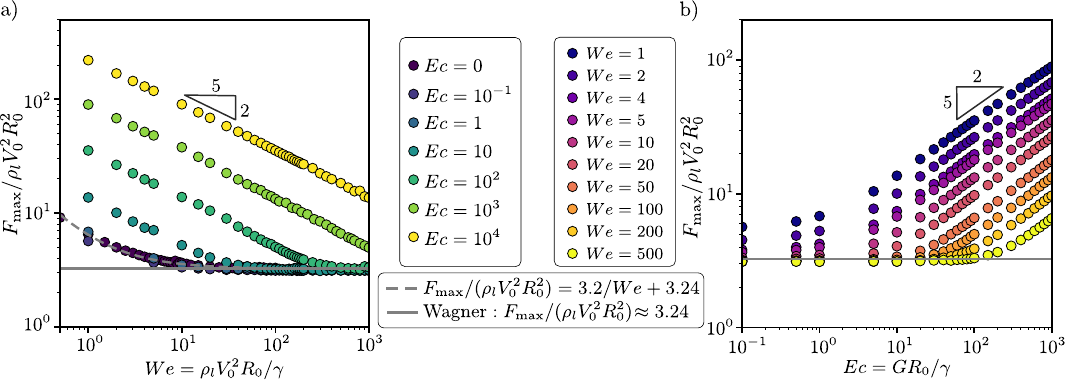}
  \caption{Peak force in the elastic-memory limit (\(De\to\infty\)): (a) Variation of the
    normalized peak force, \(F_{\max}/(\rho_l V_0^2 R_0^2)\), with the Weber number
    \(We=\rho_l V_0^2R_0/\gamma\) for different elastocapillary numbers
    \(Ec=GR_0/\gamma\). For \(Ec\lesssim\mathcal{O}(1)\) the data follow the liquid-impact
    result: a high-\(We\) Wagner plateau \(\simeq 3.24\), with the low-\(We\) correction
    \(F_{\max}/(\rho_l V_0^2 R_0^2)\approx 3.2/We+3.24\) (dashed line,
    eq.~\eqref{eq:fmax_liq}). As \(Ec\) increases, \(F_{\max}\) rises, most clearly at low
    \(We\), and for sufficiently large \(Ec\) the curves acquire a log–log slope \(-2/5\),
    i.e. \(F_{\max}/(\rho_l V_0^2 R_0^2)\sim We^{-2/5}\) at fixed \(Ec\), consistent with
    the approach to Hertzian elastic contact. (b) Dependence on \(Ec\) at fixed \(We\)
    (curves labelled by \(We\)). At small \(Ec\) all series collapse to the liquid-like
    level (\(\approx 3.24\)); above a \(We\)-dependent crossover, \(F_{\max}\) increases
  monotonically with \(Ec\), following \(F_{\max}/(\rho_l V_0^2 R_0^2)\sim Ec^{2/5}\),
again, consistent with the approach to Hertzian elastic contact. Together, (a,b) show a
continuous evolution from Wagner (liquid) to Hertz (elastic) behavior as \(Ec\)
increases.}
\label{fgr:fvswe,ec}
\end{figure*}

\begin{figure*}
  \centering
  \includegraphics[width=0.9\textwidth]{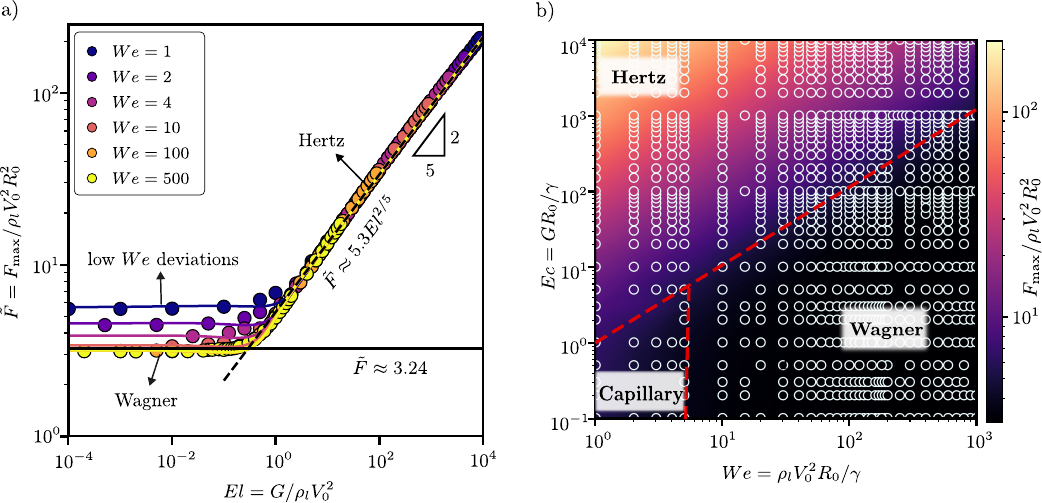}
\caption{Unified scaling and regime map:
(a) Collapse of the normalized peak force versus the elasticity number \(El=Ec/We=G/(\rho_l V_0^2)\).
For \(El\lesssim 1\) and sufficiently large \(We\) the data sit on the Wagner plateau
\(F_{\max}/(\rho_l V_0^2 R_0^2)\approx 3.24\); deviations at very small \(We\) reflect capillary
corrections in eq.~\eqref{eq:fmax_liq}. For \(El\gtrsim 1\) all cases collapse onto a single
power law with slope \(2/5\),
\(F_{\max}/(\rho_l V_0^2 R_0^2)\sim El^{2/5}\) (eq.~\eqref{eq:FmaxElscale}), the hallmark of Hertz scaling.
(b) Contours of \(F_{\max}/(\rho_l V_0^2 R_0^2)\) in the \((We,Ec)\) plane (symbols:
simulation points). The dashed guide \(El=1\) marks the smooth crossover from the
Wagner region (lower right) to the Hertz region (upper left); the low‑\(We\) corner
is capillary‑dominated and requires the correction in eq.~\eqref{eq:fmax_liq}.
The map visualizes the continuous transition from liquid‑like to solid‑like impact forces as
\(We\) and \(Ec\) are varied.}
\label{fgr:fvsel}

\end{figure*}

Collapsing the data using $El$ (figure~\ref{fgr:fvsel}a) reveals two regimes. For
$El\lesssim1$ and sufficiently large $We$, the data lie near Wagner’s constant level,
$F_{\max}/(\rho_l V_0^2 R_0^2)\approx3.24$. At small $We$, inertia competes with
capillarity and the low‑$We$ correction in eq.~\eqref{eq:fmax_liq} is required
\citep{Vatsal_thesis}. For $El\gtrsim1$, all points collapse onto a single master curve
with slope $2/5$:

\begin{equation}
\frac{F_{\max}}{\rho_l V_0^2 R_0^2}\sim El^{2/5}\sim\left(\frac{G}{\rho_l V_0^2}\right)^{2/5},
\label{eq:FmaxElscale}
\end{equation}
consistent with Hertz scaling for elastic impacts. Notably, surface tension does not enter
this high‑$El$ law, as expected for elastic solids. The transition from the Wagner
(liquid) to the Hertz (elastic) regime is smooth. A contour map over $(We,Ec)$
(figure~\ref{fgr:fvsel}b) visualizes the continuous variation of $F_{\max}$ across the
space.

\section{Theory}

In this section we identify the asymptotic limits of the peak impact force and
develop a unified predictive model. First, we derive scaling expressions for
$F_{\max}$ in the two extreme regimes -- an elastic contact limit versus a hydrodynamic
impact limit -- and then combine these results to propose a single predictive expression
for the dimensionless maximum impact force.

\subsection{Purely Elastic Limit: Hertz Contact Theory}

Consider a solid elastic sphere of radius $R_0$, mass $m$, and elastic modulus $G$ (shear
modulus, assuming an incompressible material) impacting a rigid flat
surface with speed $V_0$. Upon contact, the sphere deforms and a normal force $F$ develops
according to Hertz's contact law. For a sphere indenting a half-space, the
force–indentation relation is given by the 3/2-power law of classical Hertz contact
mechanics \citep{hertzUeberBeruhrungFester1882},

\begin{align}
  F(\delta) = m\ddot{\delta} = -\frac{4}{3}E^*\sqrt{R_0}\delta^{3/2},
  \label{eqn:F_delta}
\end{align}
where $\delta(t)$ is the indentation depth and $E^*$ is the effective Young's modulus of
the contacting pair. For a sphere against a rigid flat, $E^* = 2G/(1-\nu)$; taking
Poisson's ratio $\nu\approx0.5$ for an incompressible solid, we get $E^* \approx 4G$,
so the prefactor $(4/3)E^*\sqrt{R_0}$ in eq.~\eqref{eqn:F_delta} is about $(16/3)G\sqrt{R_0}$.

In the ideal elastic limit (no dissipation), the sphere will momentarily come to rest
at maximum compression, converting all its kinetic energy into elastic deformation energy.
Using energy conservation between the moment of impact and the instant of maximum
indentation $\delta_{\max}$ (when $\dot{\delta}=0$), we have:

\begin{align}
\frac{1}{2}mV_0^2 = \int_0^{\delta_{\max}} F(\delta)\,d\delta\,.
\end{align}

Substituting the Hertz law for $F(\delta)$ and performing the integration yields the
elastic energy stored at indentation $\delta_{\max}$:

\begin{align}
  \frac{1}{2}mV_0^2 &= \int_0^{\delta_{\max}}
  \frac{16}{3}G\sqrt{R_0}\delta^{3/2}d\delta\\
                    &= \frac{16}{3}G\sqrt{R_0}\cdot\frac{2}{5}\delta_{\max}^{5/2}\\
                    &= \frac{32}{15}G\sqrt{R_0}\delta_{\max}^{5/2}.
\end{align}

Rearranging this result to solve for the peak indentation $\delta_{\max}$ gives:

\begin{equation}
  \delta_{\max} = \left(\frac{15\,mV_0^2}{64\,G\sqrt{R_0}}\right)^{2/5}.
\end{equation}

The maximum force occurs at $\delta = \delta_{\max}$. Substituting the expression for
$\delta_{\max}$ in eq.~\ref{eqn:F_delta}, we get

\begin{equation}
  F_{\max} = \frac{16}{3}G\sqrt{R_0}\left(\frac{15\,mV_0^2}{64\,G\sqrt{R_0}}\right)^{3/5}.
\end{equation}

Normalizing with the inertial force scale $\rho_lV_0^2R_0^2$, we get

\begin{equation}
  \frac{F_{\max}}{\rho_lV_0^2R_0^2} =
  \frac{16}{3}\left(\frac{5\pi}{16}\right)^{3/5}\left(\frac{G}{\rho_lV_0^2}\right)^{2/5}
  \approx 5.3\,El^{2/5},
\end{equation}
which is consistent with our large $El$ results, cf. figure ~\ref{fgr:fvsel}a.

\subsection{Purely Liquid Limit: Wagner Impact Theory}

At the opposite extreme ($Ec \to 0$) the sphere behaves as a liquid drop, and
its impact dynamics are governed by inertia and capillarity in the classical Wagner
limit \citep{cheng2021drop,zhangImpactForcesWater2022}.
Instead of an elastic compression, the drop undergoes rapid localized deformation at the
moment of impact: the south pole flattens against the substrate while the remainder of the
drop (including the north pole) is still moving downward at nearly the impact speed. The
vertical momentum of the drop’s center of mass is redirected into a radial outflow along
the substrate, causing a small “wetted” area to grow outward from the impact
point\citep{sanjayRoleViscosityDrop2025}. This
Wagner-type mechanism -- a thin spreading lamella initiated at the contact point
\citep{Eggers2010,Josserand2016,negusModellingDropletImpact2022,negus2021droplet} --
contrasts sharply with the distributed Hertzian contact of an elastic solid. It produces a
pronounced impulsive force at touchdown, as the drop’s momentum is arrested over a short
time and small area. The first force peak thus originates from pure inertial impingement
of the liquid on the surface\citep{sanjayUnifyingTheoryScaling2025}. We have analyzed this
case in past, for details see the references \citet{zhangImpactForcesWater2022,sanjayRoleViscosityDrop2025,sanjayUnifyingTheoryScaling2025}.

During the very early stage ($t \sim \tau_\rho = D_0/V_0$), the normal force rises sharply
to its first maximum $F_{\max}$ as the drop’s inertia is transferred to the substrate.
At this moment the deformation is still localized: the contact radius has grown only to
the order of the drop’s initial radius. In fact, experiments confirm that at the peak
force time $t_{\max}$, the spread diameter $D_f(t_{\max})$ is approximately equal to the
initial drop diameter $D_0$, consistent with early-time self-similarity of the impact \citep{sanjayRoleViscosityDrop2025,sanjayUnifyingTheoryScaling2025,Mandre_etalPRL_2019,Philippi_Lagrée_Antkowiak_2016}.
Wagner’s inviscid theory predicts that the peak force scales with the inertial pressure on
the drop’s footprint. Non-dimensionalizing $F_\text{max}$ by $\rho_l V_0^2 R_0^2$ (with
$\rho_l$ the liquid density) yields a constant of order unity. Indeed, for large Weber
numbers (negligible surface tension), simulations and experiments find
$F_\text{max}/(\rho_l V_0^2 R_0^2) \approx 3.24$, cf. figure 3a of \citet{zhangImpactForcesWater2022}. At lower $We$, surface tension enhances
the impact, and the peak force is increases with decreasing $We$ (following a
$F_{\max}/(\rho_l V_0^2 R_0^2) \sim We^{-1}$ correction in this regime). This
initial peak is inertia-dominated and is relatively insensitive to liquid viscosity:
$F_\text{max}$ remains nearly constant for drops with viscosity up to about $100\times$
that of water \citep{sanjayRoleViscosityDrop2025}. 
Since our spheres have a very low background viscosity ($Oh = 10^{-2}$), in the limit $Ec \to 0$ our results for $F_\text{max}$ follow the same trends. 
Only for highly viscous drops (Ohnesorge
number $Oh \gtrsim 1$) does viscous dissipation significantly attenuate the first peak,
reflecting the fact that most of the drop’s momentum is redirected (and the force
generated) before substantial viscous effects have time to act, cf. \citet{sanjayUnifyingTheoryScaling2025}.

\subsection{Predictive Interpolating Model for Maximum Impact Force}\label{sec:combineTheTwo}

For intermediate conditions ($El \sim \mathcal{O}\left(1\right)$),
the sphere’s deformation and the fluid’s inertia both contribute, and the peak force deviates from
either pure Hertz or Wagner scaling alone. We therefore express the dimensionless peak
force as a weighted transition between the two asymptotic contributions following the
approach of Sanjay \& Lohse (2025)\citep{sanjayUnifyingTheoryScaling2025},

\begin{equation}
  \frac{F_{\max}}{\rho_lV_0^2R_0^2} = 5.3\,\text{f}\left(El\right)\,El^{2/5} +
  (1-\text{f}\left(El\right))\Big(\frac{3.2}{We} + 3.24\Big),
  \label{eqn:WagnerHertz_transition}
\end{equation}
with a smooth transition function $\text{f}\left(El\right)$ 
based on tanh function defined as
\begin{equation}
  \text{f}(El) = \frac{1+ \text{tanh}(\frac{El-a}{b})}{2},
  \label{eqn:tanh}
\end{equation}
where the $We$-dependant parameters $a \approx 1.3$ for $We \to 1$ and $\approx 0.5$ for $We \gtrsim 10$ and $b \approx 1.4$ for $We \to 1$ and $\approx 0.2$ for $We \gtrsim 10$ (see \cite{basiliskJana2026} for details of this fit). 
Here, $a$ measures the critical $El$ at transition from Wagner's to Hertz's scaling (fig.\ref{fgr:fvsel}). The width of this transition is indicated by $b$.
The remaining coefficients in eq. (\ref{eqn:WagnerHertz_transition}) are fixed a
priori: the Hertz prefactor 5.3 follows directly from the elastic analysis above
(without fitting), while the constants 3.24 (Wagner plateau) and 3.2 (the $We^{-1}$
capillary correction) are taken from the Newtonian impact model of Sanjay \& Lohse
(2025)\citep{sanjayUnifyingTheoryScaling2025}. This construction ensures a
continuous interpolation between the Hertz and Wagner limits. In spirit it follows the
additive scaling approach of Sanjay \& Lohse (2025)\citep{sanjayUnifyingTheoryScaling2025}
for drop impacts, but unlike their model -- which includes separate viscous regimes --
here only the two primary regimes (elastic vs. inertial) are needed.
The resulting formula smoothly bridges the two
asymptotes and correctly reproduces the peak-force scaling in both limits (this predictive
curve is plotted in fig.~\ref{fgr:fvsel}a for comparison).

\section{Influence of Elastic Stress Relaxation} \label{finiteDe}

\begin{figure*}
  \centering
  \includegraphics{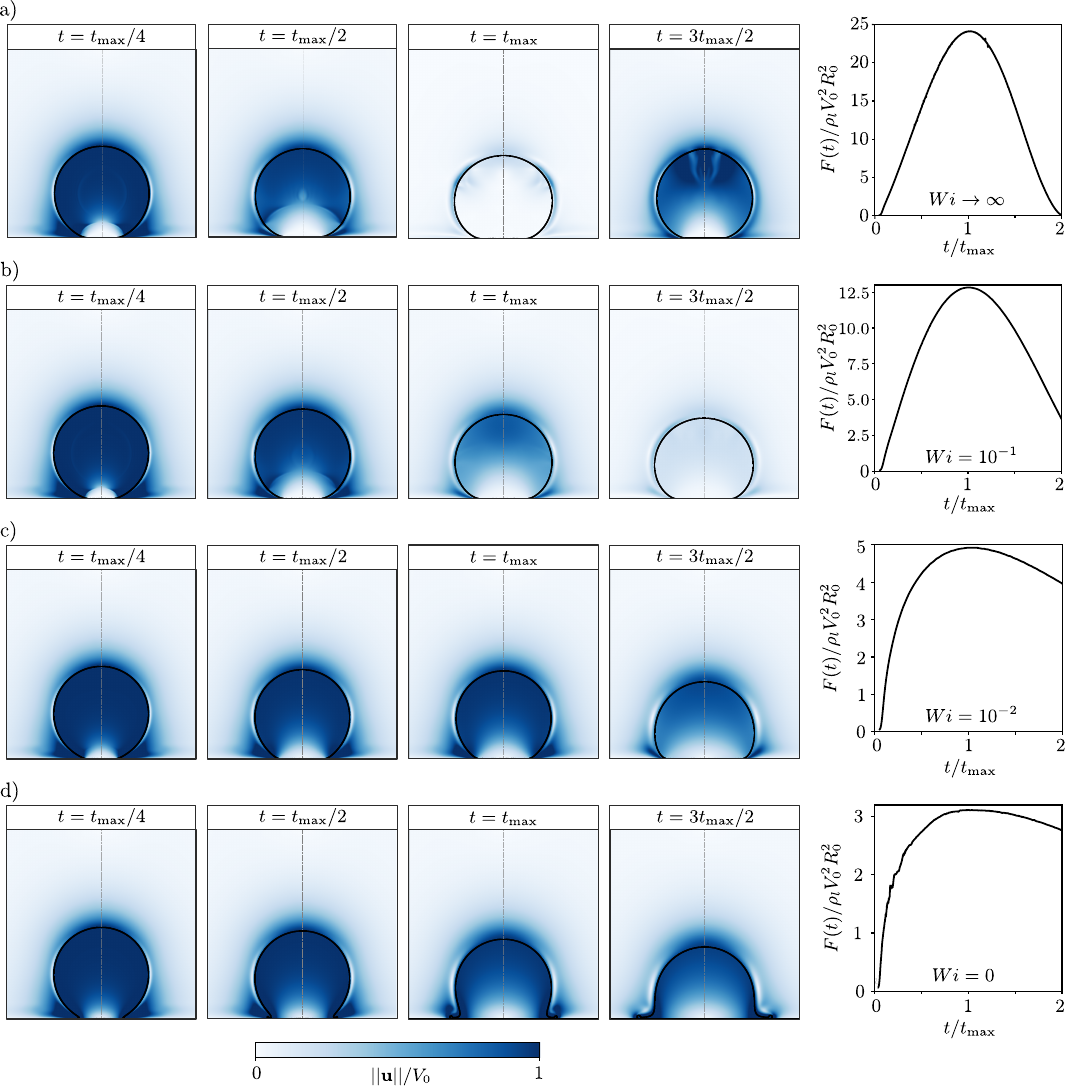}
  \caption{Relaxing material memory at fixed $We$ and $El$. Evolution of shape (left) and
  force (right) when $Wi$ decreases from $\infty$ to $0$ at $We=100$ and $El=40$: (a)
$Wi\to\infty$ (elastic‑memory limit); (b) $Wi=10^{-1}$; (c) $Wi=10^{-2}$; (d) $Wi=0$
(Newtonian).  For each case, the color scheme of each snapshot represents the
  magnitude of the velocity normalized by the impact velocity, alongside the corresponding
force history $F(t)/(\rho_l V_0^2 R_0^2)$ plotted versus $t/t_{\max}$ (right).
As $Wi$ decreases, contact time increases and $F(t)$ becomes increasingly
liquid‑like with a reduced $F_{\max}$.}
  \label{fgr:Wi_transition}
\end{figure*}

\begin{figure*}
\centering
\includegraphics[width=0.65\textwidth]{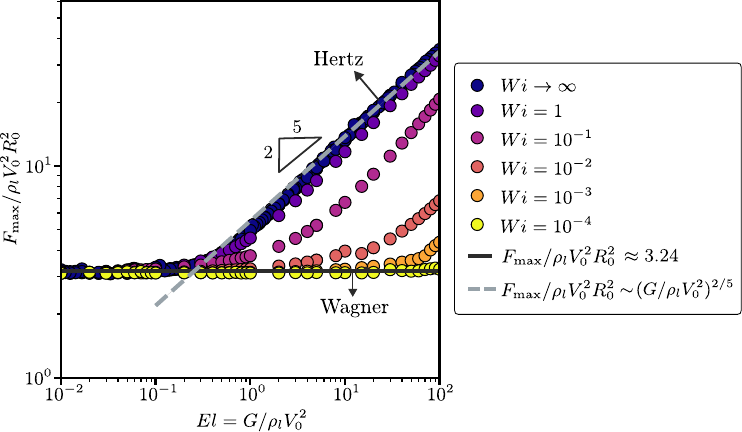}
  \caption{Peak force versus elasticity number at different $Wi$ (all at $We=100$). The
  black horizontal line indicates the Wagner plateau ($\simeq3.24$); the dashed guide has
slope $2/5$ (Hertz). Increasing $Wi$ shifts the departure from the plateau to lower $El$
and drives the curves toward the Hertz master law; at $Wi=0$ all data sit on the Wagner
level.}
  \label{fgr:fvselvarywi}
\end{figure*}

The results in \S~\ref{infDe} established the two asymptotic force scalings for impacts
with permanent memory ($De\to\infty$): a Wagner plateau at small elasticity number
and a Hertz law at large elasticity number (figure~\ref{fgr:fvsel}). These two limits also
bound impacts when the material memory is finite: relaxing the memory shifts the
response continuously from solid‑like to liquid‑like, with the transition controlled by
the non‑dimensional relaxation time. We quantify the memory effect with the Weissenberg number, $Wi$ (eq. \ref{eq:Wi}), which compares the elastic relaxation time $\lambda$ to the impact time $R_0/V_0$ (see \S~\ref{subsec:gov_eqns}). While $De$ compares $\lambda$ to the inertio‑capillary time,
$Wi$ is the more natural process‑time measure here and increases with the degree to which
elastic stresses persist during impact \citep{WiandDe_Poole}. Thus $Wi=0$ ($De=0$)
recovers a Newtonian liquid, whereas $Wi\to\infty$ ($De\to\infty$) yields an
elastic‑memory limit.

Figure~\ref{fgr:Wi_transition} (fixed $We=100$ and $El=40$) visualizes the
progressive loss of elastic behavior as $Wi$ decreases from $Wi\to\infty$ to
$0$. In the large‑$Wi$ limit, the bead contacts briefly and rebounds; the force
trace is nearly symmetric with a large peak, characteristic of Hertz‑like
loading. Reducing $Wi$ increases the contact time and skews $F(t)$ towards a
liquid‑like evolution with a much smaller peak. At $Wi=0$ the material has no
memory and behaves as a Newtonian liquid: the sphere spreads and flows, and
$F(t)$ exhibits the familiar asymmetric shape.

To quantify the role of memory, we plot $F_{\max}/(\rho_lV_0^2R_0^2)$ versus
$El=Ec/We=G/(\rho_lV_0^2)$ for several $Wi$ at $We=100$ in fig.~\ref{fgr:fvselvarywi}. At
$Wi=0$, the data follows the Newtonian liquid level established in \S~\ref{infDe}. On the
other hand, as $Wi \to \infty$, the points converge to the elastic‑memory master curve of
\S~\ref{infDe}, namely the Hertz scaling $F_{\max}/(\rho_lV_0^2R_0^2) \approx
5.3El^{2/5}$ at large $El$. Between these limits, weaker memory (smaller $Wi$) delays
the transition to Hertz scaling, while stronger memory (larger $Wi$) makes the elastic
response apparent already for softer spheres. Notably, even at $Wi\to\infty$ the plateau
persists for $El\ll1$, because the modulus is introduced exclusively through $El$: vanishing $G$
implies a liquid‑like bound irrespective of memory.

Taken together with \S~\ref{infDe}, these results show that both material
stiffness (via $El$) and material memory (via $Wi$) govern the peak impact
force: the Wagner and Hertz laws remain the bounding asymptotes, while $Wi$
sets how rapidly the system transitions between them.

\section*{Conclusions \& outlook}

In this work, we investigate the impact of a viscoelastic sphere on a non-contacting rigid
surface and chart a continuous transition in impact dynamics from liquid-like to
solid-like behavior by tuning the material parameters. The primary peak force $F_{\max}$
(associated with the inertial impact) smoothly crosses over from Wagner's inertial-drop
scaling to Hertz's elastic-contact scaling as the elasticity number $El = Ec/We$
increases. For small $El$ (liquid-like response), we reproduce $F_{\max}/(\rho_l V_0^2 R_0^2)
\to 3.24$ matching the constant plateau from Wagner's theory. In contrast, for large $El$
(elastic-dominated regime), $F_{\max}/(\rho_l V_0^2 R_0^2)$ grows following a power-law
$\approx 5.3 El^{2/5}$, consistent with Hertz's prediction for elastic spheres. These
two limiting behaviors bound the force response, and the transition between them is
gradual rather than abrupt. The Weissenberg number $Wi = De\,\sqrt{We}$, which quantifies
the polymer’s relaxation time relative to the impact time, governs this memory-driven
crossover: as $Wi$ increases from $0$ (no elastic memory) to $\infty$ (permanent memory),
the peak-force scaling shifts continuously from the Wagner limit to the Hertz limit. Thus,
by adjusting $Wi$, one can smoothly interpolate between liquid-drop and elastic-solid
impact outcomes.

The modeling choices in this work were made to isolate the physics of the
liquid–to–elastic transition and enable direct comparison to the classical limits. The
viscoelastic sphere obeys Oldroyd‑B, which neglects finite microstructure extensibility
and shear‑thinning; this is acceptable for our flow history but can be systematically
relaxed with constitutive models that incorporate finite extensibility and rate‑dependent
viscosity \citep{Snoeijer_etal_2020,Mckinley_exten_flow,Thien_1977,tanner2000}. The
substrate is non‑contacting, so the thin gas layer is present but not explicitly resolved
with lubrication and wetting dynamics; prior work shows that air cushioning, skating on a
gas film, and nanoscale first contact depend sensitively on slip, compressibility and
rarefaction
\citep{smith-2003-jfm,thoroddsen-2003-jfm,thoroddsen_etoh_takehara_ootsuka_hatsuki_2005,kolinski-2012-prl,vdveen-2012-pre,Driscoll2011,li-2015-jfm,chubynsky-2020-prl,garcia2024skating,langley2017impact,zhang-2021-prf,sprittles2024gas}.
We stress that in the present work, the substrate is assumed to remain perfectly rigid. Small elastic deformations of the substrate can, however, modify the intervening gas layer \cite{Langley2020}, which in turn may alter the impact dynamics and the measured force signature of the impacting sphere.Ma
Furthermore, the onset of contact will modify the shear stress at the interface - including effects
such as adhesion - that can modify the peak force, and will alter the stresses upon rebound.
Finally, a small background viscosity renders the ($De\to\infty$) limit to be of Kelvin–Voigt type
rather than ideally elastic; classical analyses quantify how viscoelastic dissipation and
elastic waves perturb Hertzian impact
\citep{pao1955extension,hunterEnergyAbsorbedElastic1957}. These assumptions do not alter
the governing exponents (Wagner versus Hertz) but they may affect quantitative prefactors
and very‑short‑time, near‑contact details.

Several direct extensions can sharpen and
generalize these results. Experiments with soft hydrogel/elastomer beads at moderate $V_0$
can probe the crossover regime and test the full $F(t)$ waveform (peak magnitude, rise
time, symmetry/skewness), leveraging recent studies on soft‑solid and gel impacts and
elastohydrodynamic bouncing
\citep{mitra2021bouncing,jalaal-2018-jfm,luu-2009-jfm,martouzet2021dynamic,chevy2012liquid,aguero2022impact,bertinSimilaritySolutionsElastohydrodynamic2024},
and using established force‑measurement protocols from liquid‑drop impacts
\citep{Mitchell2019,Li2014,Zhang2017,zhangImpactForcesWater2022,sanjayRoleViscosityDrop2025}.
Furthermore, incorporating a thin‑gas lubrication model with dynamic wetting (including
slip and, if needed, rarefaction/compressibility) will resolve when and how contact
initiates and how this feeds back on the very‑early‑time force
\citep{zheng2021air,smith-2003-jfm,kolinski-2012-prl,vdveen-2012-pre,li-2015-jfm,Driscoll2011,Mandre_etalPRL_2019,chubynsky-2020-prl,garcia2024skating,sprittles2024gas}.
Linking impact memory with lubricated‑impact and wetting transitions will place the
near‑contact force history on firmer ground
\citep{snoeijer-2013-arfm,Richard2000,sharma-2021-jfm,sanjay_chantelot_lohse_2023,chantelot_lohse_2021}.

\section*{Author contributions}
V.S., J.K., and D.L. conceived the study. S.J. and V.S. planned the numerical simulations.
S.J. performed the simulations and analyzed the data. V.S. and D.L. developed the theory.
S.J. and V.S. designed the structure of the manuscript. S.J., V.S., D.L. and J.K. wrote the
manuscript. V.S. and D.L. supervised the project. All authors discussed the results
and approved the final manuscript.

\section*{Conflicts of interest}
There is no conflict of interest.

\section*{Data availability}
All codes used in this work is available as an open-source repository at
\url{https://github.com/comphy-lab/Soft-Sphere-Impacts} \cite{basiliskJana2026}.

\section*{Acknowledgements}
We would like to thank Vincent Bertin, Pierre Chantelot, Maziyar Jalaal, Andrea
Prosperetti, and Jacco Snoeijer for discussions.
This work was carried out on the national e-infrastructure of SURFsara, a subsidiary of
SURF cooperation, the collaborative ICT organization for Dutch education and research.
This work was sponsored by NWO - Domain Science for the use of supercomputer facilities.
This work was supported by NWO-Canon grant FIP-II grant. V.S. acknowledges start-up
funding from Durham University.


\balance


\bibliography{rsc} 
\bibliographystyle{rsc} 

\end{document}